\documentstyle[11pt,newpasp,twoside]{article}
\begin{document}

\title{OB Associations, Open Clusters, and the\\
       Luminosity Calibration of the Nearer Stars}
\author{Anthony G.A.\ Brown}
\affil{European Southern Observatory, Karl-Schwarzschildstra{\ss}e 2, 85748
Garching bei M\"unchen, Germany}

\begin{abstract}
In the context of the luminosity calibration of the nearer stars I discuss the
Hipparcos results on distances to nearby OB associations and open
clusters. The shortcomings and assumptions in the analyses used to derive
these results are pointed out and for the open clusters a comparison is made
with results obtained from main sequence fitting. I conclude that given the
considerable uncertainties in the latter technique there is no convincing
evidence that the Hipparcos based distances to open clusters beyond the Hyades
should not be trusted.
\end{abstract}

\section{Introduction}
Every method of distance determination that relies on knowing the brightness
of some group of standard candles ultimately depends on establishing
accurately the luminosity calibration of the stars in the Solar
vicinity. Historically, this was accomplished as follows. Starting from the
Hyades, for which an accurate geometrical distance was available, the main
sequences of several well-studied clusters were patched together in order to
obtain a luminosity calibration extending from late G to O5 stars. This
calibration was then cross-checked for the early type stars with the distance
of Sco~OB2, which was also established geometrically albeit less accurately
than for the Hyades. Subsequent calibrations of, e.g., O-stars and Cepheids
were based on an extension of this main sequence fitting technique to more
distant clusters containing these types of stars.

The major weakness in this method of constructing the distance ladder is the
assumption that age differences for clusters with lifetimes up to that of the
Hyades ($\sim 600$~Myr) do not matter and that reliable corrections for
differences in the chemical compositions of clusters can be made. Hence, an
important driver for the Hipparcos mission was to determine by geometrical
means (trigonometric parallaxes and proper motions) accurate distances to the
nearby OB associations and open clusters in order to establish a fundamental
luminosity calibration.

However, apart from the distances to the Hyades and Sco~OB2 the Hipparcos
results have not remained unchallenged. They have been especially disputed in
the case of the Pleiades and other open clusters beyond the Hyades, for which
recent main sequence fitting distances apparently cannot be reconciled with
those of Hipparcos. In the following I review the Hipparcos results for the
nearby OB associations and open clusters beyond the Hyades. In each case the
weaknesses and assumptions implicit in the analysis of the Hipparcos data are
discussed and for the open clusters a discussion on the uncertainties in the
main sequence fitting technique is included.

\section{The Nearby OB Associations}
Based on the Hipparcos Catalogue (ESA 1997) a comprehensive survey of the
stellar content of nearby ($\la 1$~kpc) OB associations in 22 fields on the
sky was undertaken by de Zeeuw et~al.\ (1999). Using a combination of proper
motions and parallaxes the members of OB associations were searched for on the
assumption that those belonging to the same association share a common space
motion apart from a small velocity dispersion. This led to the successful
identification of physical groups in 12 out of the 22 fields for which
subsequently distances were determined based on the mean Hipparcos
parallax. These distance are all systematically smaller, by about
$0.2$~magnitudes in the distance modulus, than the previously established
photometric ones.

The sources of uncertainty are: (1) deviations from the kinematical model of
uniform motion plus a small and isotropic dispersion, (2) biases occuring when
converting the parallaxes of the association members into a mean distance (see
e.g., Arenou \& Luri 1999), (3) selection biases in the Hipparcos data, (4)
erroneous sky-boundaries for the associations, and (5) the presence of
interloper field stars masquerading as association members. All these points
were addressed by de Zeeuw et~al.\ (1999) through extensive Monte Carlo
simulations and their results show that it is mainly the last item which is a
cause for concern. The number of interlopers may be 10--35\% of the number of
association members and if they are primarily foreground stars one will
underestimate the distances to the associations. Nevertheless, assuming that
the distances are not biased, de Zeeuw et~al.\ concluded that the calibration
of the HR-diagram for early type stars may have to be revised. This was
further studied by de Bruijne (1999) using the following analysis.

From basic astrometry it follows that if the tangential velocity of a cluster
is known, the proper motions give information on the parallaxes. For Sco~OB2
the relative precision of the measured proper motion ($\mu/\sigma_\mu\sim25$)
is superior to the relative precision of the parallax
($\pi/\sigma_\pi\sim7$). Hence, parallaxes derived from velocities are more
precise estimates of the true parallaxes than the observed ones. Collecting
all the Hipparcos data for a particular cluster one can simultaneously
determine the cluster centroid space motion, the internal velocity dispersion,
and the individual `kinematically improved' parallaxes for all member
stars. This kinematical modeling technique is described in detail in
Lindegren, Madsen, \& Dravins (2000) and was applied through a slightly
different implementation by de Bruijne (1999) to Sco~OB2. From the results he
derived a more accurate HR-diagram for this association which shows that
indeed the previously widely used Schmidt-Kaler (1982) calibration of the
main sequence for B-stars is too bright. The main sequence as derived by de
Bruijne (1999) corresponds more closely to the calibration presented by
Mermilliod (1981). The main uncertainties are: (1) the accuracy of the
membership list and (2) the assumption of a uniform motion plus a small,
unique and isotropic velocity dispersion. These error sources are extensively
discussed in Lindegren et~al.\ (2000) and de Bruijne (1999) who both show that
their analyses are robust.

\section{Open Clusters beyond the Hyades}
Because of the closeness of the Hyades cluster, its extent on the sky, and the
large number of members in the Hipparcos Catalogue, there is little doubt
about the accuracy of the Hipparcos distance as derived by Perryman et~al.\
(1998). Thus I will turn promptly to a discussion of the open clusters beyond
the Hyades.

One of the surprising results to come out of the analysis of the Hipparcos
data was the distance to the Pleiades cluster based on the parallaxes of its
members. Analyses by van Leeuwen (1999) and Robichon et~al.\ (1999a) lead to a
distance modulus of $5.37\pm0.07$, significantly different from the value of
$5.60\pm0.04$ derived from the most recent main sequence fitting analysis by
Pinsonneault et~al.\ (1998). A similar discrepancy was found for the Coma Ber
cluster. However the most serious concern was that the differences between the
various clusters studied with Hipparcos could not easily be reconciled with
the standard picture of main sequence locus as a function of metallicity and
helium content.

This prompted Pinsonneault et~al.\ (1998) to undertake a detailed
investigation of the main sequence fitting method from which they concluded
that the discrepancy between the Hipparcos and main sequence fitting distances
to the Pleiades is too large to be explained as due to errors in the
main sequence fitting method. Instead they propose that at least in the area
on the sky close to the cluster centre there are systematic errors in the
Hipparcos parallaxes at the $\sim 1$~mas level. Pinsonneault et~al.\ (1998)
ascribe the cause of these systematic errors to the asymmetric distribution of
the Hipparcos observations over the parallax ellipse of the Pleiades
members. Narayanan \& Gould (1998) suggested that a systematic error in the
Pleiades distance may arise due to the correlated errors in the observations
of different stars in small ($\sim1^\circ$) regions on the sky.

The explanation for the source of systematic errors suggested by Pinsonneault
et~al.\ (1998) was shown by Robichon et~al.\ (1999a) to be
incorrect. Furthermore, the presence of correlations in the observations
between different stars was anticipated before the launch of the Hipparcos
mission and they can be taken into account. This was done by both van Leeuwen
(1999) and Robichon et~al.\ (1999a) and in principle there should be no
systematic errors left. However, the assumptions are that the correlations are
well understood and calibrated, that the cluster membership list is clean,
that the cluster kinematics correspond to a single velocity vector and a small
isotropic dispersion, and that the correlations are the sole cause of the
errors. In this context it is worth noting that Lindegren et~al.\ (2000) find
evidence for a log-normal distribution of the velocity dispersion in the
Hyades and that the Pleiades are known to have an elliptical halo. Hence, the
cluster kinematics might be more complicated.

On the other hand one can investigate the error sources for main sequence
fitting and there the situation is rather murkier than in the conclusions
reached by Pinsonneault et~al.\ (1998). There are five important sources of
{\em external\/} error for this technique: (1) the choice of stellar models,
(2) the choice of atmosphere models, (3) the calibration from $(L,T_{\rm
eff})$ to, e.g., $(M_V,(B-V))$, (4) the cluster metallicity and helium
abundance, and (5) the quality and homogeneity of photometric
data. Notwithstanding their extensive analysis, what Pinsonneault et~al.\
(1998) have mainly shown is that when one settles on a particular choice for
each of these five ingredients small {\em internal\/} errors on the main
sequence fitting distances can be achieved.

Robichon et~al.\ (1999b) point out that the choice of calibrations from the
theoretical to the observational HR-diagram alone can lead to up to $0.2$
magnitude effects on the main sequence. In addition, as described in both
Pinsonneault et~al.\ (1998) and Robichon et~al.\ (1999a), the metallicities of
open clusters in the Solar neighbourhood are still not accurately known. The
measurements differ from author to author and between the photometric and
spectroscopic determinations. The best precision one can obtain within a
single homogeneous study is $0.1$ dex, which already translates to a $0.1$
magnitude effect on the main sequence locus when using Johnson $B,V$
photometry. Finally, studies by Dravins et~al.\ (1997) and de Bruijne (2000),
employing the kinematic modeling technique, have succeeded in deriving a very
high precision main sequence locus for the Hyades in the $M_V$--$(B-V)$
plane. The latter has been compared to stellar models by Castellani,
Degl'Innocenti, \& Prada Moroni (2000) and they show that even in the `safe'
colour range of $0.5$--$1.0$ in $(B-V)$ the fitting of theoretical isochrones
to the main sequence is affected by significant uncertainties.

In the light of all these uncertainties one must conclude that from main
sequence fitting there certainly is no evidence that the Hipparcos distances
to open clusters should not be trusted.

\end{document}